\documentclass[journal,comsoc,10pt,letterpaper]{IEEEtran}

\usepackage{booktabs}
\usepackage{multirow} 
\usepackage{xcolor,soul,framed} 
\usepackage{url}
\usepackage[hidelinks]{hyperref}

\colorlet{shadecolor}{yellow}
\usepackage[pdftex]{graphicx}
\graphicspath{{../pdf/}{../jpeg/}}
\DeclareGraphicsExtensions{.pdf,.jpeg,.png}

\usepackage{algorithm}
\usepackage{algorithmicx}

\usepackage[noend]{algpseudocode}
\usepackage[caption=false,font=footnotesize]{subfig}

\usepackage[cmex10]{amsmath}
\usepackage{array}
\usepackage{mdwmath}
\usepackage{mdwtab}
\usepackage{eqparbox}
\usepackage{booktabs,tabularx,makecell}
\newcolumntype{Y}{>{\raggedright\arraybackslash}X} 
\usepackage{dblfloatfix}  
\usepackage{subcaption} 
\usepackage[caption=false,font=footnotesize]{subfig} 
\usepackage{xcolor}
\newcommand{\rev}[1]{{#1}}

\usepackage{amssymb}
\allowdisplaybreaks
\usepackage{booktabs,tabularx}
\usepackage{siunitx}
\begin{document}
\bstctlcite{IEEEexample:BSTcontrol}
    \title{\rev{Edge-Side Fingerprints of Service Tiering and Quota Throttling in Starlink}}
\author{Houtianfu Wang,~\IEEEmembership{Student Member,~IEEE,}%
~Hanlin Cai,~\IEEEmembership{Student Member,~IEEE,}%
~Haofan Dong,~\IEEEmembership{Member,~IEEE,}\\%
and Ozgur B.~Akan,~\IEEEmembership{Fellow,~IEEE}%
\thanks{H. Wang, H. Cai, H. Dong, and O. B. Akan are with the Internet of Everything Group, Department of Engineering, University of Cambridge, Cambridge CB3 0FA, U.K.
(e-mail: \{hw680,hc663,hd489\}@cam.ac.uk; oba21@cam.ac.uk).}%
\thanks{O. B. Akan is also with the Center for neXt-generation Communications (CXC), Department of Electrical and Electronics Engineering, Ko\c{c} University, Istanbul 34450, Turkey.}%
}  
\maketitle


\begin{abstract}
We design and evaluate an edge-side measurement procedure for auditing service tiering and quota-based throttling in Starlink. Using a 232.8-hour
plan-hopping campaign on a UK residential terminal, we align 1 Hz terminal
telemetry with host-side probes to obtain portal-labeled traces spanning
priority, post-quota throttling, stay-active operation, and residential
service. These regimes manifest as distinct signatures in goodput, PoP
RTT, and an internal-to-user ratio \(R=C_{\mathrm{int}}/T_{\mathrm{user}}\).
We further show that high-speed \(R\) is stable over 30-minute sub-windows,
that low-rate clusters have no aligned persistent obstruction or PoP-loss
signature, and that clean high-speed dips do not move \(R\) into the
low-rate band. A lightweight rule on windowed medians separates high-speed
from low-rate operation on this trace without operator visibility.
\end{abstract}




\section{Introduction}
\label{sec:intro}

Low-Earth Orbit (LEO) constellations are an increasingly important
platform for broadband connectivity. Starlink, the largest deployed
system to date, presents a familiar retail product to end users (a
small terminal and a monthly subscription) but internally exposes
multiple service tiers, data quotas, and post-quota throttling modes.
Beyond headline downlink rates, operators can map these plans to
queueing priorities, rate shaping, and fairness policies at the point of
presence (PoP). This raises a practical question for edge observers: when user-perceived throughput collapses, can we diagnose whether the link is under policy enforcement (e.g., tiering or throttling), rather than experiencing incidental low-performance episodes (e.g., short congestion, obstruction, or path variation), especially when plan/quota metadata is unavailable or not collected in measurement pipelines? Our focus is policy observability under a controlled plan-hopping design:
we test whether portal-defined regimes admit compact edge-side fingerprints.

\rev{
Prior Starlink measurement studies have characterized access performance
from several complementary views. End-to-end measurement campaigns report
latency, throughput, loss, handover behavior, routing effects, and
geographic variation using fixed terminals, mobile terminals,
crowdsourced probes, and hitchhiking measurement platforms
\cite{michel2022first,mohan2024multifaceted,ferguson2024starlink,izhikevich2024democratizing,wu2025beneath}.
These studies establish the performance baseline of Starlink access links,
but the commercial service state is usually treated as fixed background
metadata.

A second group of work studies non-policy causes of Starlink or LEO-link
performance variation, including rain and cloud attenuation, mobility,
controlled obstruction, solar-storm effects, and link-internal adaptation
to ionospheric disturbances
\cite{laniewski2024wetlinks,ullah2025impact,ramanathan2024superstorm,liu2025starlink,wang2025environment}.
This literature is important for interpreting low-throughput or high-RTT
episodes caused by the propagation environment. Our problem is different:
we ask whether portal-defined service-tier and quota states can be
identified from edge-side measurements when the terminal and local host are
controlled.

A third line of work uses terminal telemetry, protocol traces, and
emulation tools to study the implementation and reproducibility of LEO
network behavior. These works expose useful gray-box interfaces, combine
terminal-side metrics with outside-in probes, and study routing,
fine-grained downlink dynamics, transport behavior, and trace-driven
replay
\cite{rajiullah2023carl,zhao2024lens,wang2024large,zimmermann2025better,zhao2025satpipe,tian2024ebpf}.
Building on this measurement foundation, this letter targets a different
layer of observability: how service tiering and quota-triggered throttling
appear as low-dimensional, edge-diagnosable fingerprints on the same
Starlink terminal.
}
To address this gap, we introduce a policy-aware, cross-layer measurement
methodology that drives a single terminal through different service-tier and
quota states while aligning terminal-side telemetry with host-side
probes. On a residential Starlink deployment in the UK, a \rev{232.8-hour}
plan-hopping campaign exercises stay-active, priority, post-quota
throttling, and residential service. The resulting traces reveal stable,
state-dependent fingerprints in download throughput, \rev{round-trip time
(RTT)} to the PoP, and an internal-to-user ratio
\(R = C_{\mathrm{int}}/T_{\mathrm{user}}\) that combines terminal telemetry
with host-side goodput. Here \(T_{\mathrm{user}}\) is the host-side
\rev{Transmission Control Protocol (TCP)} goodput from active throughput
tests, while \(C_{\mathrm{int}}\) is an internal downlink throughput
indicator taken from the telemetry field
\texttt{downlink\_throughput\_bps} (converted to Mbps); accordingly, \(R\)
should not be interpreted as protocol header overhead. \rev{We use portal
status only for ground-truth labeling in this study; the detector itself
relies on short windowed throughput tests and local telemetry.} We report
a single-site case study: the alignment procedure is reusable, but
numerical thresholds and values are deployment-specific.

\rev{We use service tiering to denote persistent plan-level
differentiation, such as high-speed service versus the stay-active
low-rate tier, and quota throttling to denote the transition into the
post-quota \(\sim\)1\,Mbps state after the priority quota is depleted. The
stay-active tier is a separate low-rate service state, not a
quota-triggered throttling event. This distinction is central to our
labeling and interpretation.}

The trace yields three observations. First, portal-labeled plan/quota regimes manifest as three dominant operating signatures on this trace: two low-rate regimes (stay-active and post-quota throttled) and a high-speed regime (priority/residential), with distinct PoP RTT and $R$ characteristics. On this deployment, the two low-rate regimes typically appear as stable plateaus (around 0.5 Mbps for stay-active and around 1 Mbps for post-quota throttling), but these values are empirical signatures rather than labeling criteria. Second, quota depletion is followed by a short enforcement-delay window and then a sharp drop in user goodput accompanied by a step change in $R$, aligned with the portal-reported transition into post-quota throttling on this trace. Third, the joint space of download goodput and $R$ forms well-separated clusters for these portal-defined regimes, enabling a simple threshold-based heuristic to distinguish low-rate from high-speed windows and, on this dataset, to differentiate the two low-rate regimes. Together, these results provide a compact starting point for replicable edge-side auditing of service-tier and
quota policies on LEO access links and motivate exploring such fingerprints in future constellations.

\section{Methodology}

\begin{figure}[t]
  \centering
  \includegraphics[width=1\columnwidth]{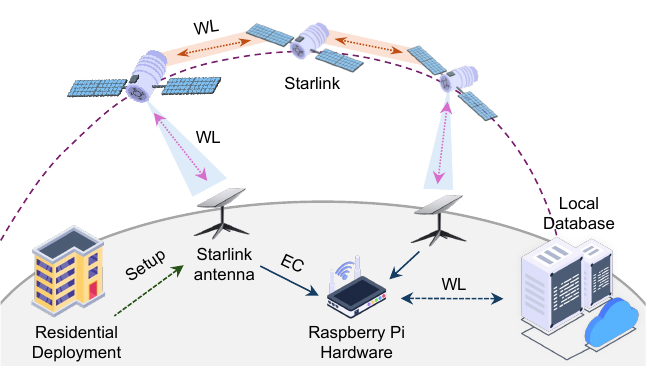}
  \caption{Measurement setup (WL: Wireless Link, EC: Ethernet Connection).}
  \vspace{-12pt}
  \label{fig:system} 
\end{figure}

\subsection{Deployment and measurement host}

As shown in Fig. \ref{fig:system}, we deploy our measurement infrastructure in a representative real-world
Starlink residential environment located at the University of Cambridge,
with open-sky view and no persistent line-of-sight obstructions. A Raspberry Pi 5 connects to the terminal via wired Ethernet and serves as a dedicated measurement host. The subscription is not shared with other clients during experiments, and all logs are timestamped in
\rev{Coordinated Universal Time (UTC) with Network Time Protocol (NTP)
clock synchronization}. \rev{The telemetry collector and the active
probing client both run on this Raspberry Pi. The collector polls the user
terminal through its local management interface; no probe is deployed
inside the Starlink infrastructure. Consequently, the internal downlink
indicator \(C_{\mathrm{int}}\), PoP RTT, loss, and obstruction flags are
terminal-reported fields, whereas the speedtest and ICMP probes measure
host-visible end-to-end performance.}

\subsection{Terminal telemetry}

We collect terminal-side telemetry using starlink-gRPC-tools%
~\cite{starlink-grpc-tools} with a 5\,s polling interval, stored in a
local SQLite database. We additionally enable bulk-history export to
obtain 1\,Hz time series including internal downlink/uplink throughput
indicators, PoP-side RTT/loss statistics, and scheduling/obstruction
flags.

\subsection{Host-side active probing}
On the same Raspberry Pi, a Python client performs end-to-end probing.
Every $\sim$4--5\,s it issues a short \rev{Internet Control Message
Protocol (ICMP)} ping train (4 probes) to a public \rev{Domain Name System
(DNS)} resolver and records average RTT and loss. \rev{Every 120 s it runs \texttt{speedtest-cli} with automatic server selection;
in its default downstream test, the tool opens multiple parallel TCP
connections to the selected server and reports their aggregate downstream
throughput. We use this aggregate downstream value as \(T_{\mathrm{user}}\),
a host-side application-layer goodput aggregated over multiple connections,
rather than a single-flow throughput or physical-layer capacity. Ethernet
byte counters are logged only as auxiliary link-activity monitors and are
not used to compute \(T_{\mathrm{user}}\).} All records use UTC timestamps to align host-side observations
with terminal telemetry.

\subsection{Policy-aware labeling via plan hopping}

We run telemetry and probing continuously for 232.8
hours total. To exercise service-tier- and quota-driven
behavior, we perform plan-hopping experiments that drive the same subscription through
stay-active, priority (pre-quota), a short post-depletion enforcement
delay window $G$, post-quota throttling, and
residential service. Each plan-to-plan switch is repeated 5 times. We
use Starlink account-portal status only for ground-truth labeling and
align portal events with the telemetry/probing logs.

We define a stable segment as a contiguous interval where the portal-reported plan/quota state is unchanged. To avoid conflating transient effects near portal transitions with steady-state regimes, we discard a guard interval of $\pm120$~s around each portal state change and label only the remaining intervals of length $\geq T_{\min}$; in our dataset, the minimum stable-segment length is approximately 1~h~18~m~50~s. The low-rate plateaus reported in Sec.~III are empirical signatures of these portal-defined regimes on this deployment, not a labeling criterion.

For each stable segment, we assign one of the nominal states in
Table~\ref{tab:states} based on the portal-reported plan/quota status.
\begin{table}[t]
\centering
\caption{\rev{Labeled service states used in the analysis.}}
\label{tab:states}
\scriptsize
{
\begin{tabular}{lll}
\hline
State & Meaning & Role \\
\hline
S1 & Stay-active low-rate plateau (\(\sim\)0.5 Mbps) & low-rate tier \\
S2 & Priority before quota depletion & high-speed \\
G  & Post-depletion before enforcement & transition / grouped high-speed \\
S3 & Post-quota state (\(\sim\)1 Mbps) & quota throttling \\
S4 & Residential service & high-speed \\
\hline
\end{tabular}
}
\end{table}
In subsequent analysis, S2 and S4 form a combined high-speed regime
(S2/S4), while S1 and S3 are the two low-rate regimes. \rev{S1 and S3 are
both low-rate (capped) states, but only S3 is quota-triggered throttling;
\(G\) is the post-depletion enforcement-delay window, grouped with
high-speed operation in the detector and analyzed separately in the quota
analysis.} On this trace, S1 and S3 typically appear as stable low-rate plateaus around 0.5 Mbps and 1 Mbps, respectively.


\section{\rev{Cross-layer fingerprints of tiering and throttling}}
\label{sec:multi_tier_properties}

This section characterizes observable differences of the same terminal
under different \rev{service-tier and quota} configurations, focusing on end-to-end
performance, terminal-side telemetry, and the dynamics around quota
depletion.

\subsection{End-to-end performance across states}

\begin{figure}[t]
\centering
\includegraphics[width=\columnwidth]{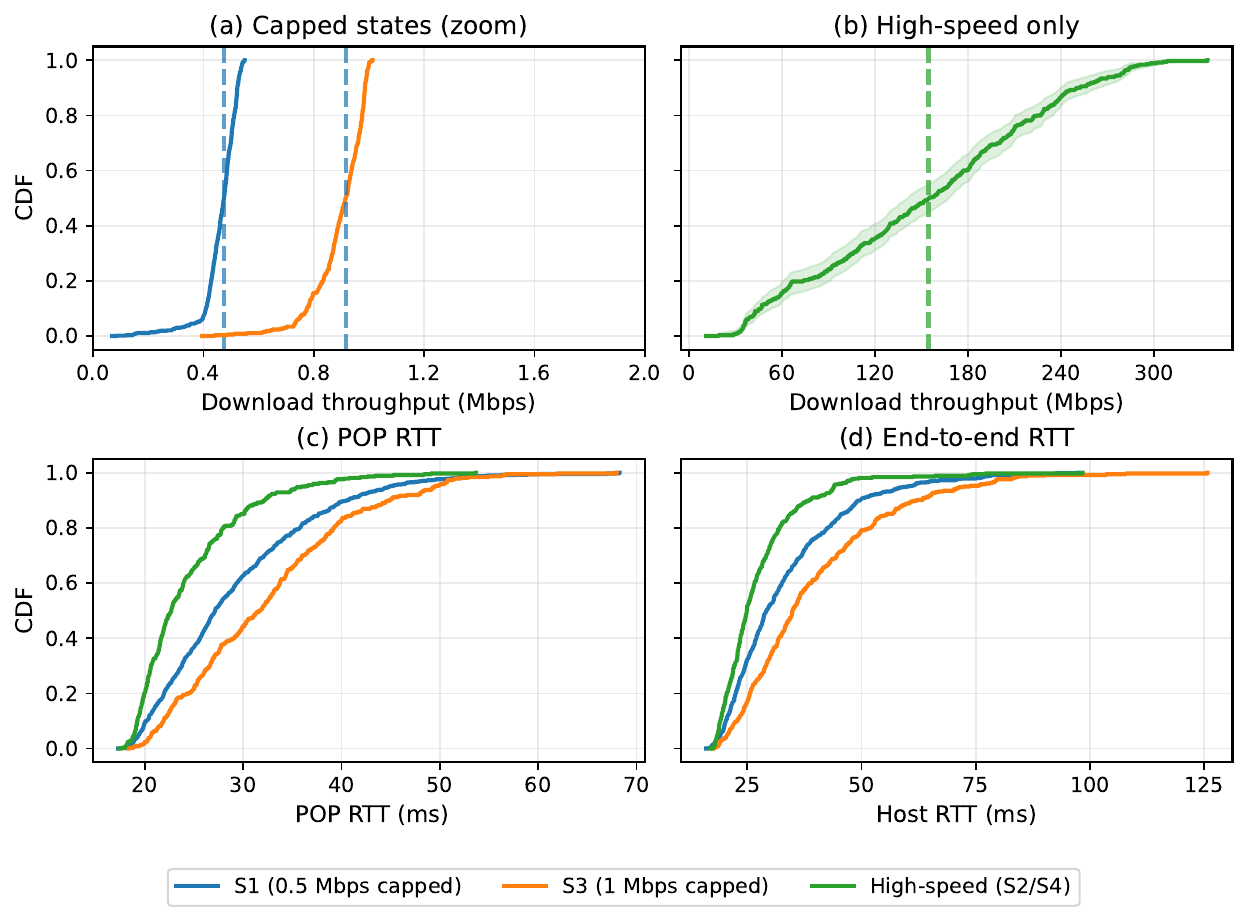}
\caption{Download throughput and RTT distributions across representative \rev{service states}.}
\label{fig:cdf}
\end{figure}

Fig.~\ref{fig:cdf} characterizes three separable operating regimes for
this terminal using download throughput and RTT distributions.
Fig.~\ref{fig:cdf}(a) focuses on the two low-rate regimes (S1 and S3):
their throughput samples concentrate tightly, forming two narrow
plateaus in the low-throughput region (centered around $\sim$0.5\,Mbps
and $\sim$1\,Mbps, respectively).
Fig.~\ref{fig:cdf}(b) shows that the high-speed regime (S2/S4) occupies
a much wider region, with most samples between a few tens and a few
hundred Mbps, clearly separated from the low-rate plateaus. Table~\ref{tab:state_summary} summarizes the per-state distributions.
S2/S4 sustains substantially higher downlink rates and the lowest median
RTT at both host and PoP.

The RTT \rev{cumulative distribution functions (CDFs)} in Fig.~\ref{fig:cdf}(c) and Fig.~\ref{fig:cdf}(d) follow
the same separation pattern: the curves for S2/S4 are shifted left,
while S1 and especially S3 extend further into the tail. Overall, for this terminal and dataset, the portal-defined regimes manifest as three dominant signatures in the throughput--latency space: two narrow low-rate plateaus and a wide high-speed cluster at the hundred-Mbps scale.

\begin{table*}[t]
\centering
\caption{Per-state summary of throughput, RTT, PoP RTT, and internal-to-user ratio.}
\label{tab:state_summary}
\scriptsize
\begingroup
\setlength{\tabcolsep}{8pt}
\begin{tabular}{lrrrr}
\hline
\textbf{State}  &
\textbf{Download Mbps (median [p10, p90])} &
\textbf{Host RTT ms (median [p10, p90])} &
\textbf{PoP RTT ms (median [p10, p90])} &
\textbf{$R$ (median [p10, p90])} \\
\hline
S1         & 0.4737 [0.4093, 0.5226] & 29.168 [20.3088, 49.4734] & 27.1594 [20.1335, 40.4985] & 21.6361 [18.9336, 25.7945] \\
S2 and S4  & 154.4406 [46.1167, 250.0403] & 24.911 [19.0640, 38.3970] & 22.6937 [19.2297, 32.0950] & 10.6677 [10.5086, 10.8386] \\
S3         & 0.9153 [0.7746, 0.9853] & 34.861 [22.6218, 61.7638] & 31.6251 [21.8994, 44.7726] & 18.0572 [16.3819, 22.0066] \\
\hline
\end{tabular}
\endgroup
\end{table*}

\subsection{Terminal telemetry and the internal-to-user ratio}

Throughput and RTT alone are often insufficient for a runtime diagnosis
of whether a low-throughput interval reflects policy enforcement or an
incidental episode (e.g., short congestion, obstruction, or shared-load
effects). We therefore incorporate terminal-side telemetry and compare a
terminal-reported internal downlink indicator against the user-visible
goodput.

\begin{figure*}[t]
\centering
\includegraphics[width=\linewidth]{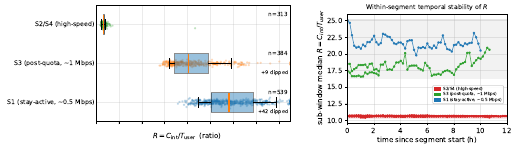}
\caption{\rev{Internal-to-user ratio \(R=C_{\mathrm{int}}/T_{\mathrm{user}}\) across service states (left; clipped at \(R=27\)) and its within-segment temporal stability (right).}}
\label{fig:ratio}
\end{figure*}

The \rev{left panel} of Fig.~\ref{fig:ratio} shows the distribution of an internal-to-user ratio
\[
R = \frac{C_{\mathrm{int}}}{T_{\mathrm{user}}},
\]
where $T_{\mathrm{user}}$ is the host-side TCP goodput from our active
throughput tests and $C_{\mathrm{int}}$ is a terminal-reported internal
downlink throughput indicator (e.g., \texttt{downlink\_throughput\_bps}).
$C_{\mathrm{int}}$ is a terminal-reported internal downlink indicator, whereas $T_{\mathrm{user}}$ is end-to-end payload goodput.

In the high-speed state S2/S4, $R$ is tightly concentrated around 10.7,
with 10th--90th percentiles between 10.5 and 10.8. In contrast, the
low-rate regimes exhibit substantially larger ratios, with a median
around 21.6 for S1 and 18.1 for S3. Together with PoP RTT statistics in
Table~\ref{tab:state_summary}, this yields two observations. \rev{First, high-speed operation exhibits a stable empirical baseline around
\(R\approx10.7\) on this dataset; we do not interpret the physical semantics
of \(C_{\mathrm{int}}\).} Second, the upward shift of $R$ in low-rate regimes makes $R$ a compact cross-layer feature for the detector in Section~\ref{sec:state_detection}.

\rev{For temporal stability, we compute 30-minute sub-window medians of \(R\)
within each stable segment (right panel of Fig. 3). High-speed windows stay
within \(\approx[10.4,10.8]\), while the low-rate windows remain separated
even with one drifting segment (\(\approx+0.2\) \(R\)/h); the minimum
high/low gap exceeds 5.8.}

\rev{Health telemetry shows no persistent obstruction episode or PoP-loss burst
aligned with the low-rate clusters: the low-rate obstruction metric is
comparable to a same-day high-speed reference, the PoP-loss 95th percentile
is zero or near zero, and the longest loss/obstruction runs last only
5--10 s. We also find that none of 20 clean high-speed throughput dips
moves into the elevated low-rate \(R\) band; 19 stay near the high-speed
baseline, and the only exception (\(R_{\mathrm{med}}=4.35\)) falls below,
not within, the low-rate band.}

\subsection{Quota depletion and enforcement-delay behavior}

\begin{figure}[t]
\centering
\includegraphics[width=\columnwidth]{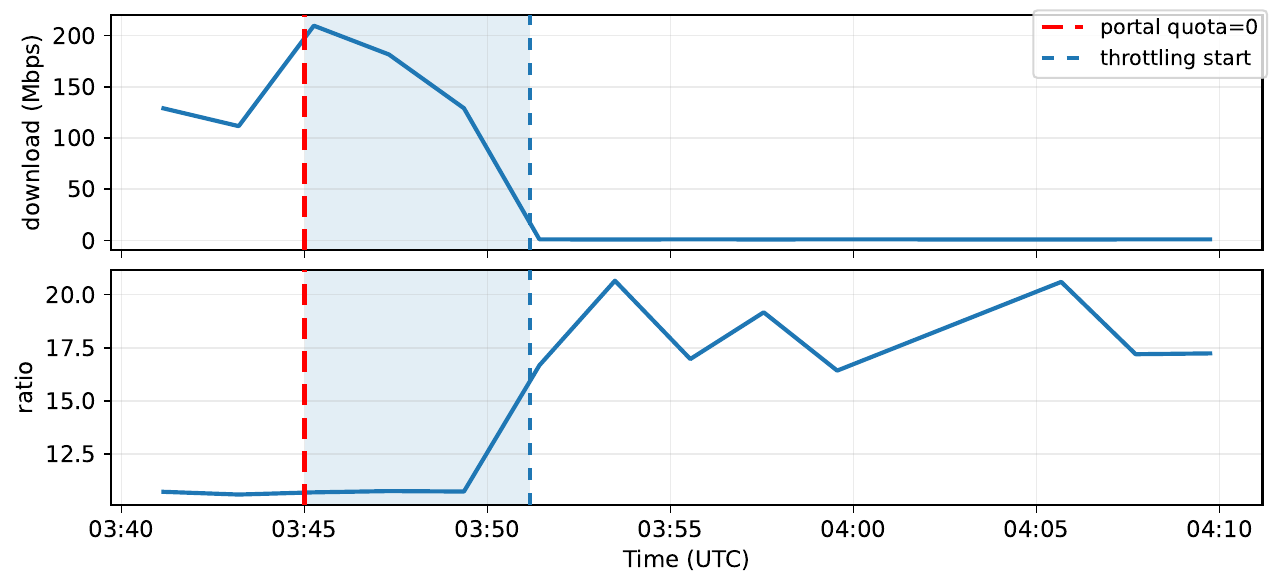}
\caption{Quota depletion, enforcement-delay window $G$, and throttling onset for the priority plan.}
\label{fig:grace}
\end{figure}

To examine the dynamics after priority quota depletion,
Fig.~\ref{fig:grace} zooms into one quota-depletion instance on
November~8, 03{:}40--04{:}10\,UTC, aligning download throughput, the
ratio $R$, and the portal's quota indicator. The red vertical line marks
the time when the portal reports zero remaining quota, the blue vertical
line marks the onset of the 1\,Mbps cap, and the shaded region
highlights the interval between the two (window $G$).

In this instance, immediately following the quota depletion report (red
line), the download throughput does not instantly collapse. Instead, it
remains in the high-speed regime for several minutes, gradually
declining from $\sim$200\,Mbps to $\sim$120\,Mbps, before abruptly
dropping to the 1\,Mbps plateau at the throttling onset (blue line).
This behavior is consistent with an enforcement delay between quota
accounting updates and scheduler-side policy activation on this trace.

The ratio $R$ provides a synchronous signal for the transition. During
both the pre-quota and enforcement-delay phases, $R$ stays at the low
values characteristic of high-speed operation. Once throttling starts,
$R$ quickly jumps to the higher plateau associated with the 1\,Mbps
low-rate regime. Along the time axis, the joint evolution of throughput
and $R$ reveals a three-stage process: quota reaches zero, followed by a
high-speed enforcement-delay window $G$, and finally a strict cap. We
report this as a case-study observation on this dataset: it indicates a
minutes-scale gap between portal quota status and throttling onset that
can be captured by lightweight active measurements.

\section{A lightweight policy-state detector}
\label{sec:state_detection}

Section~\ref{sec:multi_tier_properties} revealed structured differences
in throughput, RTT, and the internal-to-user ratio $R$ across states.
This section operationalizes these differences into an edge-side
detector: service states form separable fingerprints in a low-dimensional
feature space and can be identified with a simple rule.

We focus on three representative regimes: the low-rate stay-active state
(S1), the low-rate post-quota throttled state (S3), and the high-speed
regime (S2/S4) that covers priority before quota depletion, the
enforcement-delay window $G$, and residential high-speed operation. In
the detector, we treat $G$ as part of high-speed operation.




\subsection{Service-tier fingerprints and a decision rule}
\label{subsec:state_detection_method}

We distinguish the regimes using short active tests and local telemetry. To emulate a ``user runs a speedtest'' scenario, we partition the measurement trace into fixed-length windows of $W=180$~s with a step of 180~s (non-overlapping).

\rev{For each window, we extract the median active-test download goodput and
the median \(R\), computed by aligning each active-test timestamp to the
nearest telemetry sample within \(\pm10\) s.}

\begin{figure}[t]
  \centering
  \includegraphics[width=\columnwidth]{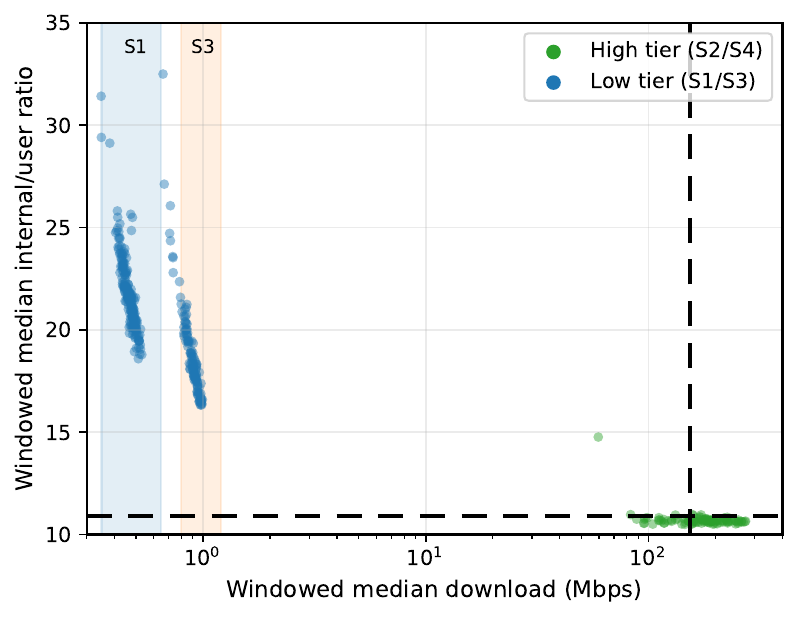}
  \caption{Windowed service-state fingerprints in the \((T_{\mathrm{user}},R)\) feature space.}
  \label{fig:state_signature}
\end{figure}

We visualize these windows in the two-dimensional feature space shown in
Fig.~\ref{fig:state_signature}. Windows from the high-speed regime
(S2/S4) cluster in a region of high throughput and low $R$, while
windows from the low-rate regimes (S1/S3) cluster in a region of low
throughput and high $R$.


\rev{The separation is wide enough for a static rule: a window is high-speed
only if its median download goodput exceeds \(T_d=50\) Mbps and its median
ratio satisfies \(R<T_r=14.5\). Table II reports per-sample statistics,
whereas the detector uses windowed medians, so these thresholds act as
conservative deployment-specific guardrails.}


\rev{On the labeled windows, this rule separates S2/S4 from S1/S3 without
errors on this trace. The two low-rate states are then distinguished by
their empirical plateaus, around 0.5 Mbps for S1 and 1 Mbps for S3.}


\section{Discussion and conclusion}
\label{sec:discussion}

Previous sections introduced an edge-side, fingerprint-based
policy-state detector. We now summarize the limitations of this single-terminal study and conclude.



\subsection{Limitations and transferability}

\rev{This is a single-site, single-terminal case study over 232.8 hours and a
finite set of service-tier/quota configurations; the numerical values and
thresholds are deployment-specific. The trace does not include adverse-weather or sustained-obstruction events; validating the same separator under those conditions requires broader deployments.
The contribution is methodological: a controlled plan-hopping template that
aligns portal state, terminal telemetry, and host-side probes to recover
low-dimensional service-state fingerprints. Applying the same procedure
across additional terminals, locations, firmware versions, and LEO systems
is left for future work.}

\subsection{Conclusion}

Using one residential Starlink terminal and a policy-aware plan-hopping
campaign, we examined how service-tier and quota
configurations manifest in edge-observable metrics on a live LEO access link. By aligning terminal
telemetry with host-side probes, we identified three stable operating
regimes and showed that a two-dimensional feature space spanned by
download goodput and $R$ yields an approximately linearly separable
fingerprint on this dataset. \rev{In particular, quota depletion exhibits a dynamic
signature---an enforcement-delay window \(G\) followed by a step into the
\(\sim\)1\,Mbps post-quota state with a synchronous jump in
\(R\)---distinguishing quota throttling from the persistent stay-active
tier.} Compared to prior Starlink measurements
focused on coverage, topology, or environmental factors, our results add
a policy-observability dimension and distill a minimal edge-side recipe:
drive the terminal through available states, collect cross-layer
telemetry, and recover low-dimensional signatures for each regime. While
the thresholds are deployment-specific, the procedure provides a compact
starting point for replicable edge-side auditing across LEO access
links.

\bibliographystyle{IEEEtran}
\bibliography{references}



\end{document}